\documentclass
[superscriptaddress,secnumarabic,amssymb,amsmath,nobibnotes,aps,prd,showkeys,showpacs,twocolumn,nofootinbib]{revtex4}%
\usepackage{bm}
\usepackage{hyperref}
\usepackage{mathrsfs}
\usepackage{xcolor,color,graphicx,graphics}
\usepackage[all]{xy}
\usepackage{epsfig}
\usepackage{latexsym,amssymb,amsmath,amsfonts} %\usepackage{amstex}
\usepackage[english]{babel} %, spanish, portuguese
\usepackage[OT1]{fontenc}
\usepackage[latin1]{inputenc}
\usepackage{makeidx}
\usepackage{hyperref}
\usepackage{color,graphicx,graphics,wrapfig,epsf}%,psfig
\usepackage{caption}
\usepackage{mwe}
\usepackage{float}
\usepackage{subfloat}
\usepackage{subfig}
\hypersetup{urlcolor=BlueViolet,
	    citecolor=Plum,
	    linkcolor=PineGreen}
\begin{document}

\title{Interiors of terrestrial planets in metric-affine gravity}

\author{Aleksander Kozak}
\email{aleksander.kozak@uwr.edu.pl}
\affiliation{Institute of Theoretical physics, University of Wroclaw, pl. Maxa Borna 9, 50-206 Wroclaw, Poland
}

\author{Aneta Wojnar}
\email{aneta.magdalena.wojnar@ut.ee}
\affiliation{Laboratory of Theoretical Physics, Institute of Physics, University of Tartu,
W. Ostwaldi 1, 50411 Tartu, Estonia
}

\begin{abstract}
    Using a semi-empirical approach we show that modified gravity affects the internal properties of terrestrial planets, such as their physical characteristics of a core, mantle, and core-mantle boundary. We also apply these findings for modeling a two-layers exoplanet in Palatini $f(R)$ gravity.

\end{abstract}

\maketitle

\section{Introduction}

Discoveries of exoplanets in Milky Way \cite{first,catalogue} and in the Whirlpool Galaxy \cite{whirlpool}, as well as growing observational data set of those objects provided by the current and future missions \cite{webb,nancy,tess,spitzer,nn} have increased a need of theoretical tools which allow to describe the planets' interiors and eventual habitable properties on the base of those data. A common approach is to extrapolate the  Preliminary Reference Earth Model (PREM) \cite{prem} and its later improvements \cite{kustowski,iasp91,aki135} (see more at \cite{iris}). Therefore, although the Earth-like planet should have at least six differently composed layers, one usually considers two \cite{seager}, iron core and silicate mantle, since they have the biggest impact on the observed properties, such as planets' mass, radius, and polar moment of inertia. However, a very different composition of the rocky planets may also be possible, as argued in \cite{quartz}, such as quartz-rich mantles, in comparison to the Solar System ones, whose mantles are mainly made of silicates. Clearly, such findings call for more research in planetary physics, not only from observational point of view, but also theoretical modelling. 

Regarding the planet's modelling based on PREM, we are still improving our knowledge on the deepest zones of the Earth, as well as the instrumentation and methods used are getting ameliorated, allowing to get more accurate data of the planet's interior. For instance, the recent seismic observation \cite{mush} has revealed the existence of the liquid/mushy region of the inner core instead of the solid one, as it has been believed to be so far.
On the other hand, a new generation of the neutrinos' telescopes will be settled to provide information on the matter density inside the planet, and on characteristics and abundances of light elements in the outer core \cite{topography,top,top2,top3}. Also in laboratories, with the use of lasers \cite{laser}, the high pressures and temperatures, that is, the extreme conditions of the Earth's core, are recreated in order to understand the properties and behaviour of iron, which is the main element of planets' cores. All those revelations make the research regarding planets' modelling relevant, especially agreeing with the fact that various models of gravity predict different layers' structure in comparison to the Newtonian model \cite{olek3}, commonly used in the planetary science. Therefore, knowing the planet's profile with the high accuracy, that is, the number of differently composed layers and their thickness, might be another tool to test theories of gravity (see the details on the method in \cite{olek2}).

As already mentioned, some extensions of Einstein theory of gravity may impact the internal structure of the rocky planets, as well as their properties \cite{olek3,olek2,aneta_jup}. This is so since such theories modify the non-relativistic hydrostatic equilibrium equations \cite{review}, and others, crucial for stellar and planetary modelling. For instance, in the Schwarzschild criterion, which is used to constitute a type of the energy transport through an astrophysical object, there appears additional term making the star more or less stable with respect to convective processes \cite{aneta_hay}; energy production in
a stellar core is also affected \cite{sak,hyd,lit,ros}, as well as stars' evolution \cite{ms}, or cooling processes of substellar objects \cite{maria}. Therefore, modified gravity theories proposed to provide some explanations of dark matter and dark energy phenomena \cite{Copeland:2006wr,Nojiri:2006ri,nojiri2,nojiri3,Capozziello:2007ec,Carroll:2004de}, space-time singularities \cite{Senovilla:2014gza}, extreme masses of compact objects \cite{lina,as,craw,NSBH,abotHBH,sak3}, or to unify all four interactions into a single theory  \cite{ParTom,BirDav}, also impact modelling of gravitational systems for which full relativistic description is not necessary. 

One of such a theory we are interested in is a subclass of the so-called Ricci-based theories \cite{ricci}, that is, Palatini $f(R)$ gravity. The main geometric property of these theories is that the metric and connection are considered as independent objects in comparison to most extensions of Einstein's one. However, their most important feature is related to their vacuum dynamics, since it provides the same dynamical equations as General Relativity ones with a cosmological constant, providing that those proposals pass Solar System test \cite{junior} and gravitational waves' observations as the waves are moving in those theories with the speed of light. But the difference is clear when one deals with matter fields - Ricci-based gravities introduce then terms which depend on energy density, modifying the structural equations \cite{gda}.

In this work we will focus on a gravitational model which introduces a quadratic Ricci scalar term, and it will be considered in the Palatini approach. Since those terms contribute to the structural equations of spherical-symmetric low-temperature spheres, such a modification will have an influence on internal properties of the planet. Therefore, using an analytical method allowing to obtain the core and core-mantle boundary values of pressure from given masses and radii of transiting exoplanets, we will demonstrate that those values will differ in modified gravity. Moreover, we will also use them to model an exoplanet interior.

\section{Simple model of small rocky planets in Palatini gravity}

In this section we will recall the hydrostatic equilibrium equations for a cold low-mass spherical symmetric object. Our terrestrial planets, with masses from the range $M_p\in(0.1 - 10) M_\oplus$, where $M_\oplus$ is the Earth's mass, and core mass fraction (CMF), defined as
\begin{equation}
    \text{CMF}=\frac{M_\text{core}}{M_p},
\end{equation}
not exceeding $\sim0.4$ of the total planet's mass\footnote{The exoplanets of Mercury's type, having cores with masses $\sim0.7$ of the total mass \cite{seager}, are excluded from such an analysis.}, will be modelled as a two-layer planet, that is, consisting of an iron core and a silicate mantle. Then, using the semi-empirical expression relating the CMF with the radius and mass of a transiting exoplanet, we will derive the planet's internal characteristics, such as core's pressure and density, their boundary values between the core and mantle, and the mantle's ones.

\subsection{Planets' structure equations }

Non-relativistic hydrostatic equilibrium equations for the quadratic Starobinski model\footnote{For full relativistic equations in Palatini gravity, see \cite{aneta_stab,olek}.}
\begin{equation}\label{star}
    f(\mathcal{R})=\mathcal{R}+\alpha\mathcal{R}^2
\end{equation}
considered in Palatini approach are given by \cite{gda,aneta_pol}
\begin{align}
    p'(r)&= -\frac{Gm \rho}{r^2}\left(1-\beta\kappa^2(5\rho-2r\rho')\right),\label{press}\\
    m(r) &=    \int^{r}_0 4\pi\tilde{r}^2\rho(\tilde r) d\tilde r,\label{mass}
\end{align}
where prime denotes the derivative with respect to the radial coordinate. Let us notice that the different numerical factors appearing in the modification term in (\ref{press}) are the results of the considered assumptions; for example, in \cite{aneta_pol} the equations were obtained by assuming the conformal invariance of the standard polytropic equation of state for the quadratic model demonstrated in \cite{fatibene}, while the equations derived in \cite{gda} are more general, without adopting any equation of state. In this work we also use some polytropic equations of state, however it differs slightly with respect to the common one, as explained in the further part of the text.

Our small rocky exoplanet is modelled as a cold sphere consisting of two different layers. As already mentioned, the material they are made of is iron in the core and silicate in the mantle, whose equations of state are given by the Birch equation of state \cite{birch,poi}, working well when temperatures can be considered uniform but less than $2000$K, and when pressure is below
$200$ GPa.
However, in order to be able to consider more massive objects than the terrestrial planets of the Solar System, one has to take into account the electron degeneracy, as the internal pressure can be $p\gtrsim10^4\,\text{GPa}$. The usual procedure is to match this equation of state with the Thomas-Fermi-Dirac one \cite{thomas, fermi, dirac, feynmann} which also qualifies to describe density-dependent correlation energy \cite{sal} which appears because of the interactions between electrons when they obey the Pauli exclusion principle and move in the Coulomb field of the nuclei. Such a hybrid equation of state is very well approximated by a modified polytropic equation of state (SKHM) of the form \cite{seager} 
\begin{equation}\label{pol}
    \rho(p)=\rho_0 +cp^n,
\end{equation}
whose best-fit parameters $\rho_0$, $c$, and $n$ for iron and silicate (Mg, Fe)SiO$_3$ are provided in the table \ref{tabpoly}. Because solids and liquids are incompressible at the low pressure regimes, the additional term $\rho_0$ is present to include this effect. Such a constructed equation of state is valid up to $p=10^{7}\,\text{GPa}$, giving the maximal value of the central pressure possible in our analysis.
\begin{table}
\caption{Best-fit parameters for the SKHM equation of state (\ref{pol}) obtained in the reference \cite{seager}.}
\centering
\begin{tabular}{llll }
\hline\noalign{\smallskip}
Material & $\rho_0$ (\text{kg m}$^{-3}$) & $c$ (\text{kg m}$^{-3}$ \text{Pa}$^{-n})$ & $n$ \\
\noalign{\smallskip}\hline\noalign{\smallskip}
Fe($\alpha$) & 8300 & 0.00349 & 0.528\\
 (Mg, Fe)SiO$_3$ & 4260 & 0.00127 & 0.549  \\
\noalign{\smallskip}\hline
\end{tabular}\label{tabpoly}
\end{table}

Moreover, to explore the model with the described features, one needs to establish the initial and boundary conditions. In the previous works we have used the shooting method in order to find the initial values of the core's densities as well as between the layers' ones \cite{olek2,olek3}. It demonstrated that modified gravity can have a significant impact on those values and this is a result of different physical assumptions such as for example Newtonian physics. Therefore, even slight modification to the standard hydrostatic equilibrium equation will have an effect on the internal structure. Having this in mind, we have restudied a simple but reasonable method \cite{zeng2} used to obtain the internal characteristic of a distant planet, whose mass and radius can be found by the use of the transit observation techniques \cite{nasaexo}.
Therefore, for the given total mass of the planet and its radius we will derive the central pressure, its value on the core-mantle boundary (CMB), and the mantle one. It will show that modified gravity indeed affects them.

\subsection{Internal structure of Palatini planets}

There is only one planet whose interior structure and materials, that is, equations of state, are known: the Earth\footnote{However, we will be equipped with the Mars ones, too, thanks to the Seismic Experiment for Interior Structure from NASA's MARS InSight Mission's
seismometer \cite{nasa}.}. The many-layers structure, their thickness, and equations of state are given by seismic data, that is, PREM \cite{prem}. Since some planets of our Solar System and exoplanets are alike dense and possess similar other characteristics, one usually extrapolates the Earth's model to describe them. Therefore, extrapolating the Earth's model, one may derive the semi-empirical expression for the core mass fraction (CMF) which carries the information on the core-mantle boundary, often used in numerical procedures and simulations of very distant planets, whose mass $M_p$ and radius $R_p$ are given by the transit. Such a relation between CMF and observed radius and mass was given in \cite{zeng1}:
\begin{equation}\label{cmf}
    \text{CMF}=\frac{1}{0.21}\left[ 1.07- \left(\frac{R_p}{R_\oplus}\right)/\left(\frac{M_p}{M_\oplus}\right)^{0.27} \right],
\end{equation}
where $R_\oplus$ and $M_\oplus$ are the Earth's radius and mass, respectively. Furthermore, CMF can be also used to obtain the approaximated value for the core radius fraction (CRF), defined as
\begin{equation}\label{crf}
      \text{CRF}=\frac{R_\text{core}}{R_p},
\end{equation}
which is also suitable for numerical analysis \cite{zeng2}:
\begin{equation}\label{sqrt}
    \text{CRF}\approx\sqrt\text{{CMF}}.
\end{equation}
Using these two values, that is, CMF and CRF, we will derive the core's and mantle's pressure, as well as its boundary value, for an exoplanet of the mass $M_p$ and radius $R_p$.

Let us firstly use the definition of local gravity, defined as usually:
\begin{equation}\label{local}
    g=\frac{Gm(r)}{r^2}, 
\end{equation}
to rewrite the equation (\ref{press}) in a more suitable form for further purposes: 
\begin{equation}
    p'(r)= -g\rho\left(1-\beta\kappa^2 
    \left[\frac{14g+g'r-2g''r^2}{4\pi Gr}\right]
    \right).
\end{equation}
Using the mass equation (\ref{mass}) together with the expression for the local gravity (\ref{local}) it can be tranformed into
\begin{equation}
    \frac{dp}{dm}=-\frac{g^2}{4\pi G}  \frac{d\, \text{ln}(m)}{dm}\sigma,
\end{equation}
where $\sigma=1-\beta\kappa^2 \left[\frac{14g+g'r-2g''r^2}{4\pi Gr}\right]$ while $\text{ln}(m)$ is the natural logarithm of $m$. Assuming that the surface pressure is zero, we integrate the above equation from the surface inward, such that
\begin{equation}\label{press2}
    \int^\text{interior}_\text{surface}dp=
    -\frac{1}{4\pi G}\int^\text{mass enclosed inside}_{M_p}g^2 d\text{ln}(m) \sigma.
\end{equation}
Before going further, let us define the surface gravity $g_s$ as a local gravity on the planet's surface with mass $M_p$ and radius $R_p$
\begin{equation}
       g_s:=\frac{GM_p}{R_p^2},
\end{equation}
while the so-called typical pressure $p_\text{typ}$ is defined as
\begin{equation}
    p_\text{typ}:=\frac{g_s^2}{4\pi G}=\frac{GM_p^2}{4\pi R_p^4}.
\end{equation}

Since the local gravity of the mantle can be assumed to be a constant \cite{zeng2}, we may integrate (\ref{press2}) to get the pressure of the mantle:
\begin{align}\label{pmantle}
    p_\text{mantle}&=2p_\text{typ}\text{ln}\left(\frac{R_p}{r}\right)\nonumber\\
&\times \left[
    1+\beta\kappa^2 \frac{7g_s}{\pi G}\frac{M_p}{R_p}\left(\frac{1}{\sqrt{M_p}}-\frac{1}{\sqrt{m}}\right)
    \right],
\end{align}
where we have used the planet's characteristics defined before. In particular, the pressure on the core-mass boundary (CMB) can be obtained by inserting $r\rightarrow R_\text{core}$ and $m\rightarrow M_\text{core}$ such that
\begin{align}\label{cmb}
    p_\text{CMB}&=p_\text{typ}\text{ln}\left(\frac{1}{\text{CMF}}\right)\nonumber\\
&\times \left[
    1+\beta\kappa^2 \frac{7g_s\sqrt{M_p}}{\pi GR_p}\left(1-\frac{1}{\sqrt{\text{CMF}}}\right)
    \right],
\end{align}
where we have used (\ref{crf}) and (\ref{sqrt}).

On the other hand, since in our model the core density $\rho_\text{core}$ can be assumed to be a constant value, the core mass is given as $M_\text{core}=\frac{4}{3}\pi R^3_\text{core}\rho_\text{core}$. Therefore, the hydrostatic equilibrium equation (\ref{press}) can be written with the use of (\ref{mass}) as 
\begin{align}
    \frac{d p_\text{core}}{dr}&=-g\rho_\text{core}\left(1-\beta c^2\kappa^2\left[\frac{9m'}{4\pi r^2}-\frac{m''}{2\pi r}\right]\right)\nonumber\\
        &=-\frac{3r p_\text{typ}}{R^2_\text{core}}\left[ 1-\beta c^2\kappa^2\frac{15g_s}{4\pi G R_\text{core}} \right].
\end{align}

Integrating the above equation results as
\begin{equation}\label{pcore}
    p_\text{core}(r)=p_0-\frac{3}{2}p_\text{typ} \left(\frac{r}{R_\text{core}}\right)^2\left(1 - 15\beta \kappa^2\frac{g_s}{4\pi G R_\text{core}}\right),
    \end{equation}
where $p_0$  is the central pressure which can be determined by matching the above $p_\text{core}$ at CMB with the pressure on the boundary (\ref{cmb}):
\begin{align}
    p_0&=p_\text{CMB}+\frac{3}{2}p_\text{typ}\left( 1-15\beta \kappa^2\frac{g_s}{4\pi G R_\text{core}} \right) \nonumber\\
   & =p_\text{typ} \left( \frac{3}{2}\left[1-15\beta \kappa^2\frac{g_s}{4\pi G R_\text{core}}\right]\right.\\
  &\left.+ \text{ln}\left(\frac{1}{\text{CMF}}\right) \left[
    1+\beta \kappa^2 \frac{7g_s\sqrt{M_p}}{\pi GR_p}\left(1-\frac{1}{\sqrt{\text{CMF}}}\right)
    \right]
   \right).\nonumber
\end{align}
The above result allows to find an approximated value of the central pressure for a given terrestrial exoplanet whose mass and radius are provided by the transit observations. The effect of modified gravity is clearly present, therefore in the next section we will numerically solve the structural equations with the use of those findings.

\section{Numerical solutions}

\begin{figure*}
\subfloat[K2-36 b, $M = 3.9M_\oplus, R = 1.43R_\oplus$]{\includegraphics[width = 3.5in]{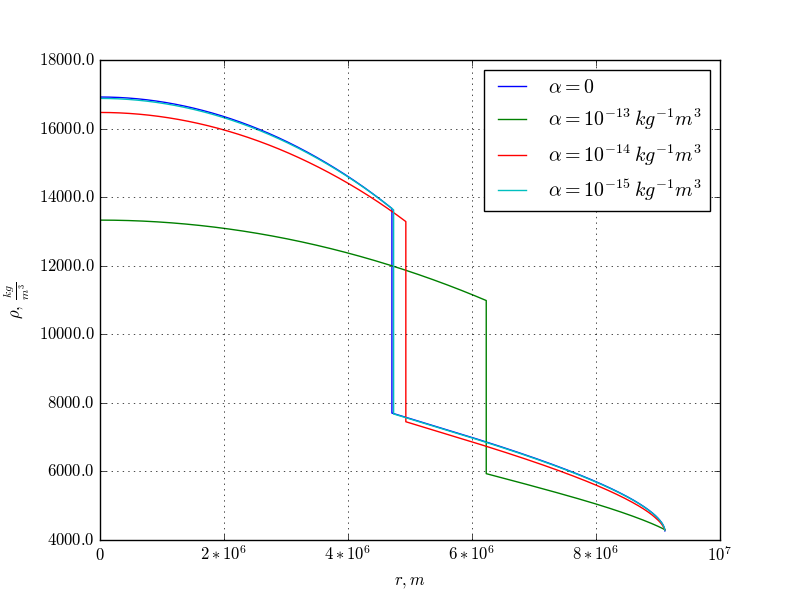}} 
\subfloat[Kepler-10 b, $M = 4.6M_\oplus, R = 1.48R_\oplus$]{\includegraphics[width = 3.5in]{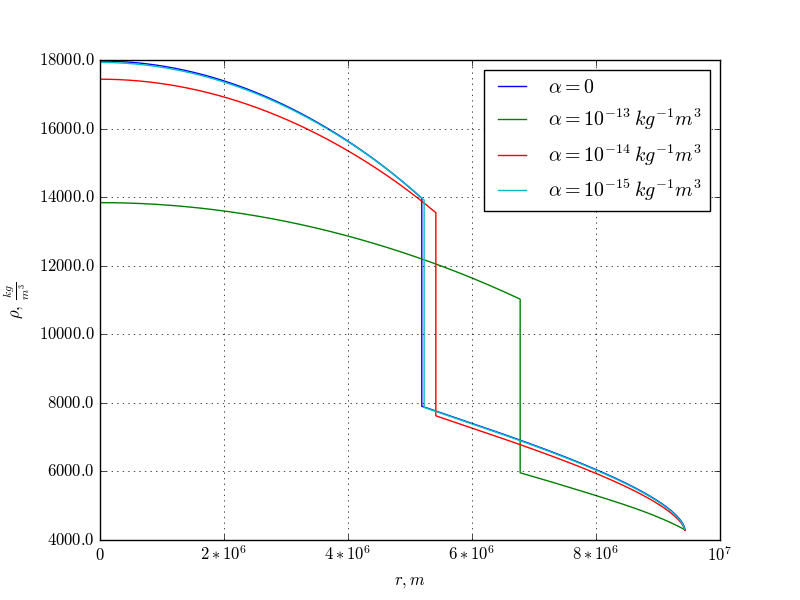}}\\
\subfloat[Kepler-20 b, $M = 9.7M_\oplus, R = 1.87R_\oplus$]{\includegraphics[width = 3.5in]{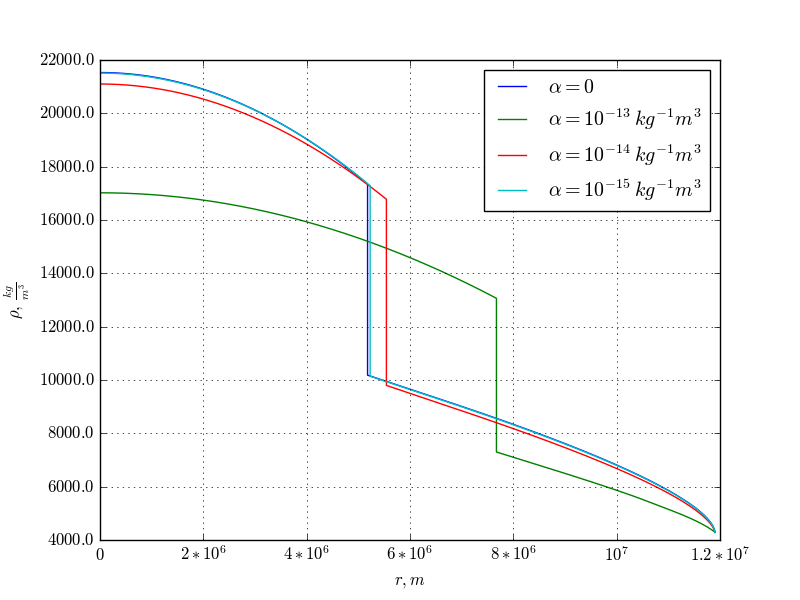}}
\subfloat[Kepler-78 b, $M = 1.97M_\oplus, R = 1.12R_\oplus$]{\includegraphics[width = 3.5in]{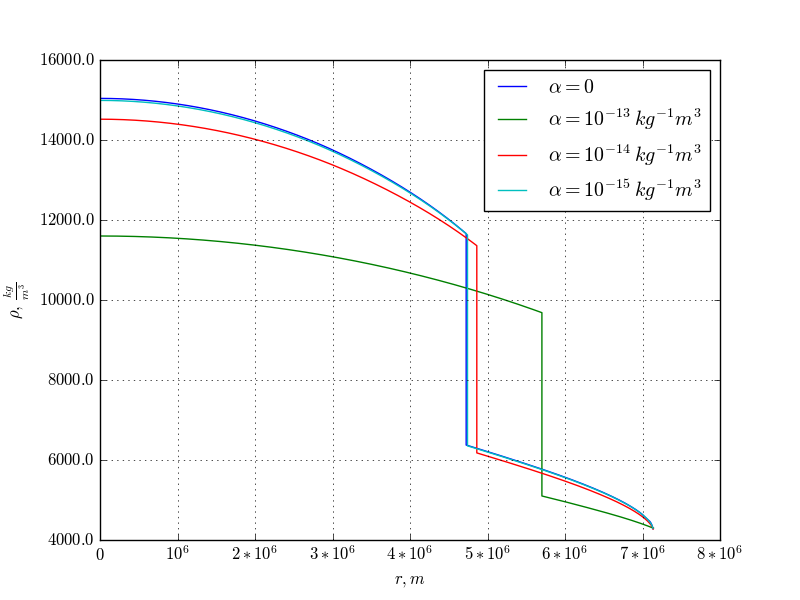}} 
\caption{[color online] Density profiles for four different Earth-like exoplanets, for different values of the parameter $\alpha = c^2\kappa^2 \beta$. The planets are assumed to be composed of two layers: iron core and mantle made of (Fe, Mg)SiO$_3$.}
\label{figures1}
\end{figure*}

Before comparing  models of different values of the Starobinsky parameter $\beta$, we introduce a dimension-full parameter $\alpha = c^2 \kappa^2 \beta$, which allows one to write the formulas in a more convenient way. We chose four values of the parameter: $\alpha\in \{0, 10^{-15}, 10^{-14}, 10^{-13}\}$, neglecting the possibility of negative values of the parameter\footnote{We do so in order to avoid reaching nonphysical solutions being a fact of the conformal transformation, for which there exists a singular value of $\alpha<0$. To learn more about that feature, see \cite{aneta_pol,olek}.}. Having established the range of the parameter, we aim at solving numerically the equations (\ref{press}) and \eqref{mass}, supplemented with the equation of state \eqref{pol}. The fact that the masses and radii of the planets we examine are fixed by the transit observations provides the possibility to determine the core density and core size with its mass, as well as to plot the density profiles. As one can see in Figure \ref{figures1}, all curves denoting solutions for different values of $\alpha$ end at the same point; what changes is the size of the core. This allowed us to compare CMFs and CRFs obtained from the quasi-empirical formula \eqref{cmf} (which is constant once the mass and the radius of the planet are given) to the numerical findings. 

\begin{figure*}[h]
\centering
    \includegraphics[width=0.7\linewidth]{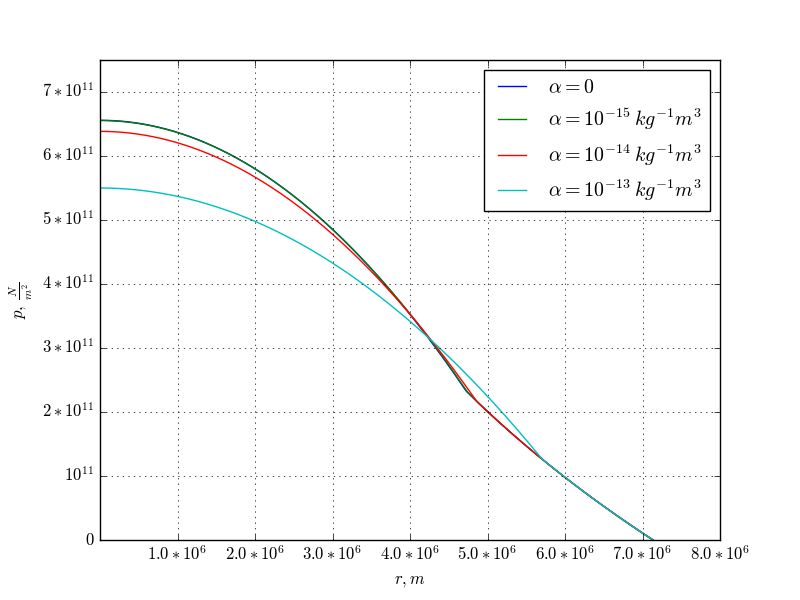}
\caption{[color online] Relation between pressure and radius for Kepler-78 b exoplanet calculated analytically using the formulas derived in this work. The curves are plotted for four different values of the parameter $\alpha = c^2\kappa^2 \beta$. The planet is assumed to be composed of two layers.}
\label{fig2}
\end{figure*}

\begin{figure*}[h]
\centering
    \includegraphics[width=0.7\linewidth]{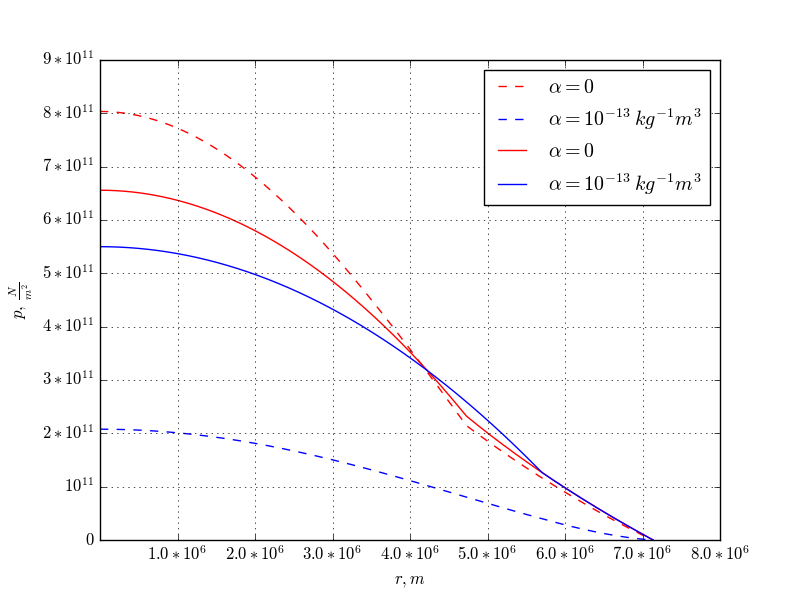}
\caption{[color online] Relation between pressure and radius for Kepler-78 b exoplanet calculated analytically and numerically. The dashed line represents the numerical solution, whereas the solid line - analytical one. The curves are plotted for two different values of the parameter $\alpha = c^2\kappa^2 \beta$. }
\label{fig3}
\end{figure*}

As far as the pressure is concerned, we simply calculate it for one planet, Kepler-78 b\footnote{But the results are similar for the other ones, too, with the more significant differences for larger planet's masses with respect to the Newtonian solutions.}, using the formulas \eqref{pmantle} and \eqref{pcore}, as well as the exact value of $R_\text{core}$ determined in the previous, numerical step. The results are shown in Figure \ref{fig2}, illustrating the effects of modified gravity on pressure within the exoplanet. The analytical solutions are then compared with numerical ones, to determine how good the approximations are. The results are shown in Figure \ref{fig3} for two values of $\alpha$. 

\section{Conclusions}
Previous studies regarding terrestrial planets in modified gravity \cite{olek3} revealed that extensions of Einstein's gravity alter the internal structure of those objects, providing a possibility to test such theories with the use of seismic data \cite{olek2}. Therefore, the physical quantities, such as core pressure and energy density, as well as their boundary values between layers, should also be affected, which would have an impact on the way we model distant planets, where seismology cannot be applied. This fact forces us to look for methods allowing to find those values, when only the observed characteristics, such as, for example, mass and radius of a transiting exoplanet, are available. In this work we wanted to check if such methods are model-independent.

As clearly demonstrated, the methods can indeed depend on the applied theory of gravity. For this analysis, we have considered quadratic modification to the General Relativity's Lagrangian (\ref{star}) considered in Palatini approach, however our conclusions are valid for other theories of gravity which modify the non-relativistic limit of their field equations. \begin{itemize}
    \item Density profiles, as already noticed in our previous works, can significantly differ in modified gravity with respect to the Newtonian model. We observe not only lower values of central density and on the core-mantle boundary, but also the cores of the given exoplanets are bigger; that is, the cores are less dense in the case of Palatini gravity. Therefore, the observed transiting planets can have different structure for the same masses and radii than the one predicted in the usual way, and can affect the planet's polar moment of inertia. 
    \item Similar situation happens when we plot the pressure curves obtained in this work: its central values decrease in modified gravity, however when we approach the planet's surface, the mantles do not differ much. This result comes from the fact that the additional term in the equation \eqref{pmantle} for the pressure in the mantle is small, and smaller that the extra term appearing in the analogous equation for the core \eqref{pcore}.
    \item We have also compared the numerical solutions for the pressure obtained from the equations (\ref{press}) and \eqref{mass} to the ones resulting from the analytical approach (which are approximated solutions). One notices that in the case of Newtonian gravity ($\alpha=0$) analytical (approximated) solution tends to provide smaller values than the numerical one. However, in the case of modified gravity, the effect is reverse - approximated analytical solution provides larger values than the numerical one. This can be explained in the following way: the analytical approximation does not take into account the effect of modification of gravity in the CMF formula \eqref{cmf}, so it stays constant for various values of the parameter $\alpha$ (since it depends of the mass and radius of the planet only, and these values do not change). On the other hand, the numerical method suggests that size of core and its mass grow in modified gravity, and thence the CMF must change. This combined effect of change in $\alpha$ and CMF/CRF results in bigger drop in internal pressure.
    \item Moreover, as already mentioned in previous point, our numerical analysis revealed that the equation for the semi-empirical CMF used in that work also depends on modified gravity. It is not a surprise remembering the fact that for finding that relation, one uses PREM model which is based on Newtonian gravity. 
\end{itemize}
Although our studies presented in this paper are based on crude methods and assumptions, such as spherical-symmetric, non-rotating planets, their two-layers structure and constant values for the mantle's characteristics, it is evident that alternative theories of gravity do impact the planets' descriptions and modelling. Improving our analytical and numerical methods, that is, taking into account the missing ingredients mainly related to more realistic planet's geometry should also manifest similar results. The work along these lines is currently underway.

\vspace{5mm}
\noindent \textbf{Acknowledgement.} 
This work was supported by the EU through the European Regional Development Fund CoE program TK133 "The Dark Side of the Universe". 
AK is a beneficiary of the Dora Plus Program, organized by the University of Tartu.

\end{document}